\documentclass[letterpaper]{article} 
\usepackage{aaai25}  
\usepackage{times}  
\usepackage{helvet}  
\usepackage{courier}  
\usepackage[hyphens]{url}  
\usepackage{graphicx} 
\usepackage{natbib}  
\usepackage{caption} 
\usepackage{algorithm}
\usepackage{algorithmic}
\usepackage{xcolor}

\usepackage{booktabs}

\usepackage{newfloat}
\usepackage{listings}
\DeclareCaptionStyle{ruled}{labelfont=normalfont,labelsep=colon,strut=off} 
\floatstyle{ruled}
\newfloat{listing}{tb}{lst}{}
\floatname{listing}{Listing}

\title{Computational Reproducibility of R Code Supplements on OSF}
\author{
    Lorraine Saju,
    Tobias Holtdirk,\\
    Meetkumar Pravinbhai Mangroliya,
    Arnim Bleier
}
\affiliations{
    GESIS – Leibniz Institute for the Social Sciences\\
    Cologne, Germany\\
    \{lorraine.saju, tobias.holtdirk, meetkumar.mangroliya, arnim.bleier\}@gesis.org
}

\begin{document}

\maketitle

\begin{abstract}
Computational reproducibility is fundamental to scientific research, yet many published code supplements lack the necessary documentation to recreate their computational environments. While researchers increasingly share code alongside publications, the actual reproducibility of these materials remains poorly understood. 

In this work, we assess the computational reproducibility of 296 R projects using the \textit{StatCodeSearch} dataset. Of these, only 264 were still retrievable, and 98.8\% lacked formal dependency descriptions required for successful execution. To address this, we developed an automated pipeline that reconstructs computational environments directly from project source code. Applying this pipeline, we executed the R scripts within custom Docker containers and found that 25.87\% completed successfully without error. 

We conducted a detailed analysis of execution failures, identifying reproducibility barriers such as undeclared dependencies, invalid file paths, and system-level issues. Our findings show that automated dependency inference and containerisation can support scalable verification of computational reproducibility and help identify practical obstacles to code reuse and transparency in scientific research.
\end{abstract}

\begin{figure}[!t]
    \centering
    \includegraphics[width=0.7\linewidth]{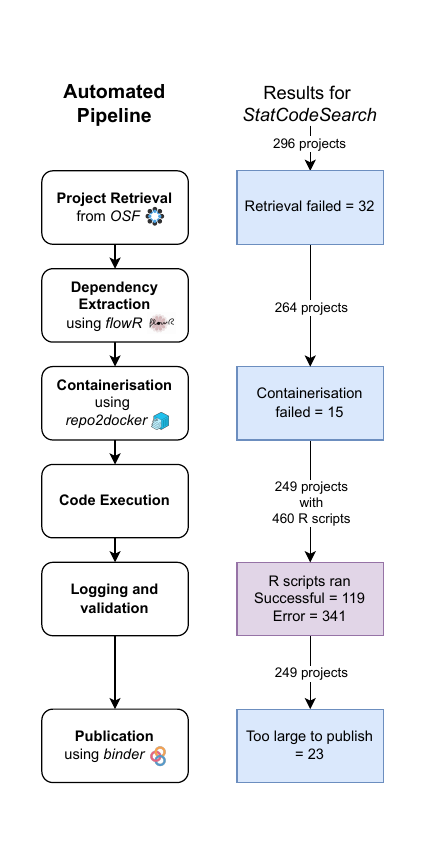}
    \caption{Our completely automated pipeline \texttt{osf-to-binder} (left) and the analysis results using this pipeline for the \textit{StatCodeSearch} dataset (right).}
    \label{fig:flow}
\end{figure}

\section{Introduction}

Ensuring computational reproducibility is increasingly recognized as a cornerstone of credible scientific research \cite{peng2011, seibold2021, NAP25303}. Several works, beginning in the 1990s \cite{claerbout1992}, have highlighted the importance of making research outputs reproducible and have proposed computational methodologies and best practices to achieve this goal. 

A fundamental aspect in achieving computational reproducibility is the provision of all necessary materials, including \textit{the raw data and the code} used to generate the results. \citet{barba_terminologies_2018} paraphrases \citet{claerbout1992} with what is now widely considered the ideal of computationally reproducible research. 
\begin{quote}
An article about computational science in a scientific publication is not the scholarship itself, it is merely advertising of the scholarship. The actual scholarship is the complete software development environment and the complete set of instructions which generated the figures.
\end{quote}
Following this goal, \citet{chung-hong__chan_what_2024}, among others, emphasise that sharing of all code and data is crucial for transparency and forms the basis for computational reproducibility. To facilitate the execution of their shared code, researchers must pay close attention to documenting their \textit{computational environment} thoroughly. This includes explicitly listing all software dependencies and their versions, using files like \texttt{requirements.txt} or including detailed information about their used environment using commands such as \texttt{sessionInfo()} in R. Containerisation technologies such as Docker, which use a \texttt{Dockerfile} to create consistent and isolated environments across different systems (e.g., different operating systems), are another widely accepted approach in the literature on computational reproducibility \cite{boettiger2015}. Furthermore, \citet{schoch2024} emphasise that external dependencies such as online \textit{APIs} are also part of the potentially changing environment, which can undermine computational reproducibility. Achieving reproducibility requires a multifaceted, proactive approach that includes transparent sharing of materials and thorough documentation of the computational environment by the authors. \citet{sandve2013} and \citet{kohrs2023} condense these requirements into basic rules for reproducible computational research.

Achieving computational reproducibility should not be assumed; it requires external verification. \citet{hardwicke2020} manually examined 250 articles from the social science literature and found that fewer than 3\% made their analysis scripts available. \citet{Rainey_Roe_Wang_Zhou_2025} found that only about 12\% of quantitative research articles provided access to both the data and the code. \citet{trisovic2022} executed R code from replication datasets hosted on the Harvard Dataverse repository in a clean runtime environment and found that 74\% of R files failed to complete without error. \citet{pimentel2019} and \citet{samuel2024} examined the reproducibility of Jupyter notebooks, mostly written in Python, and found that only 24\% and 11.6\% respectively ran without errors in a fully automated analysis. \citet{chung-hong__chan_what_2024} tested the reproducibility of 30 papers and found that, even after manual restoration of the code, at least 20\% were only partially reproducible. Furthermore, both practitioners \cite{lasser2020,nust2021} and guides \cite{arnold2019,bleier2025c} emphasise the role of services like MyBinder in enabling authors to share their analysis scripts in a way that allows for easy verification by others. 

In this work, we empirically test the computational reproducibility of 296 R code supplements published as projects on the Open Science Framework (OSF) repository. However, unlike earlier approaches \cite{trisovic2022} that used a clean runtime or manual intervention \cite{chung-hong__chan_what_2024} to establish reproducibility, we apply an automated approach to infer and extract dependencies that are necessary for a successful execution. Our work is guided by the following research questions: At what rate are we able to verify the computational reproducibility of the submissions published on OSF? Can automatic dependency inference aid in successful re-execution? Can a statistical analysis of replication failure modes inform recommendations on best practices for the publication of code supplements, and if so, what are these best practices?

\section{Methodology}

The starting point for this study was the \textbf{StatCodeSearch dataset}, which is part of the GenCodeSearchNet benchmark suite \cite{diera_gencodesearchnet_2023}. This dataset, available on HuggingFace\footnote{\url{https://huggingface.co/datasets/drndr/statcodesearch}}, consists of code-comment pairs extracted from R scripts hosted on the Open Science Framework (OSF)\footnote{\url{https://osf.io}}. It focuses specifically on R projects in the social sciences and psychology, particularly those involving statistical analysis.

The dataset contains 1,070 code-comment pairs drawn from 558 unique R scripts across 296 distinct OSF projects. While the dataset is organized at the level of individual code-comment pairs, our goal is to reconstruct interactive, reproducible computational environments at the project level.

To achieve this, we used the project identifiers provided in the dataset to retrieve the corresponding research materials. We then employed the OSFClient API\footnote{\url{https://github.com/osfclient/osfclient}} to download the full contents of each associated OSF repository.

An initial verification step revealed that, out of the 558 R code files referenced across 296 OSF projects in the \texttt{StatCodeSearch} dataset, 63 files from 32 distinct projects were no longer accessible through their original OSF directories. This outcome suggests that a portion of the dataset had become outdated, likely due to file deletions, renaming, or changes to project access permissions on the OSF platform following the initial data collection.

While OSF supports the creation of immutable, time-stamped project snapshots through its registration feature, our analysis found that only 58 out of 296 projects had used registrations, and only 49 of those preserved the files referenced in the dataset. Moreover, registered snapshots are not automatically created or mandatory, and their selective use makes it difficult to systematically recover the original state of all materials. The lack of widespread adoption of OSF registrations and the absence of robust version control systems (such as those provided by Git) make it challenging to replicate the computational environment used in these studies at the time of publication. 

Following the identification and removal of unresolvable file references from the initial dataset, the remaining 264 projects were examined for files that could support the reproduction of the original analyses. The downloaded project contents were systematically searched for reproducibility-relevant files that document the computational R environment, including \texttt{renv.lock}, \texttt{sessionInfo.txt}, \texttt{sessionInfo.RData}, \texttt{.Rprofile}, \texttt{DESCRIPTION}, \texttt{dependencies.R}, \texttt{dependency.R}, \texttt{Dockerfile}, \texttt{environment.yml}, and \texttt{install.R}.

To enable the automated execution and validation of project files associated with the GenCodeSearchNet dataset, we developed an automated pipeline, \texttt{osf-to-binder}\footnote{\url{https://github.com/Code-Inspect/osf-to-binder}}, which is publicly available on GitHub. The goal of this pipeline is to generate verifiably reproducible computational environments directly from the source code of scientific publications hosted on OSF.

The \texttt{osf-to-binder} pipeline operates through the following steps (see Figure \ref{fig:flow}):

\begin{itemize}
    \item \textbf{Project Retrieval}: 
    Given one or more OSF project identifiers, the pipeline automatically downloads and unpacks the entire file storage associated with each project.
    \item \textbf{Dependency Extraction}: For projects containing R scripts, the pipeline employs \texttt{flowR} \cite{sihler_flowr_2024}, a static dataflow analyser and program slicer to automatically extract dependencies. 
    \item \textbf{Docker Configuration}: The extracted R dependencies are used to generate a \texttt{DESCRIPTION} file, an R package metadata file that is essential for specifying dependencies in Docker-based environments. 
    \item \textbf{Containerisation}: Using \texttt{repo2docker} \cite{forde2018}, the pipeline builds a Docker container based on the project directory. It scans the repository for standard configuration files (e.g., \texttt{DESCRIPTION}) and creates a runnable Docker image accordingly. 
    \item \textbf{Code Execution}:  Within the built container, the pipeline executes all identified R scripts in a fully isolated and dependency-managed environment.
    \item \textbf{Logging and open validation}: Execution results and logs are recorded to ensure transparency and support both internal and external validation.
    \item \textbf{Publication}: To support open reproducibility, the resulting Docker image is published to a container registry (DockerHub)\footnote{All Docker images are published at \url{https://hub.docker.com/u/meet261}.}, and the project code is made available via a version control system (GitHub)\footnote{All GitHub repositories are maintained under \url{https://github.com/code-inspect-binder}.}. Additionally, we generate a MyBinder \cite{ragan2018} launch link, enabling users to run the environment in a remote RStudio instance without any local setup.
\end{itemize}

By automating dependency extraction, environment configuration, containerisation, and execution, the \texttt{osf-to-binder} pipeline offers a scalable and transparent approach to enhancing computational reproducibility for OSF-hosted research projects. It thereby supports broader efforts toward open and verifiable science.

\section{Results}

An analysis of the remaining 264 OSF projects, identified after excluding those for which the referenced R scripts could not be located during initial verification, revealed a limited presence of files commonly associated with computational reproducibility. The results are detailed in Table \ref{tab:reproducibility_files}.

\begin{table}
    \centering
    \caption{Presence of reproducibility-related files in the 264 analysed OSF projects.}
    \begin{tabular}{lrr}
        \toprule
        Dependency File & \multicolumn{2}{c}{Projects} \\
        \cmidrule(lr){2-3}
        & Total & Percentage \\
        \midrule
        \texttt{DESCRIPTION} & 2 & 0.8\% \\
        \texttt{Dockerfile} & 1 & 0.4\% \\
        \texttt{renv.lock} & 0 & 0.0\% \\
        \texttt{sessionInfo.txt} & 0 & 0.0\% \\
        \texttt{sessionInfo.RData} & 0 & 0.0\% \\
        \texttt{.Rprofile} & 0 & 0.0\% \\
        \texttt{dependencies.R} & 0 & 0.0\% \\
        \texttt{dependency.R} & 0 & 0.0\% \\
        \texttt{environment.yml} & 0 & 0.0\% \\
        \texttt{install.R} & 0 & 0.0\% \\
        No dependency file & 261 & 98.8\% \\
        \bottomrule
    \end{tabular}
    \label{tab:reproducibility_files}
\end{table}

These findings highlight the current state of explicit reproducibility provisions within the examined subset of OSF R Projects. The scarcity of these files suggests that many projects may lack readily available instructions or specifications for recreating computational environments. 

\subsection{Containerisation success and failures}
Of the 264 projects processed by the pipeline, containerisation failed for 15, comprising 35 R scripts (around 5\% of the total). The main reasons included:

\begin{itemize}
    \item \textbf{Malformed or incomplete \texttt{DESCRIPTION} files:}  Generated from \texttt{flowR} outputs, these often lacked required fields or had formatting errors, rendering them invalid.
    
    \item \textbf{Incorrectly extracted dependencies:} In several cases, \texttt{flowR} misidentified variables, file paths, or internal objects as package names, resulting in invalid entries such as \texttt{unknown}, \texttt{NULL}, or numeric values like \texttt{'0'}, \texttt{'1'}.
    
    \item \textbf{Invalid scoped package references:} Calls like \texttt{knitr::opts\_chunk} were mistakenly treated as standalone packages, which are not installable.
    
    \item \textbf{Unavailable or incompatible packages:} Some projects listed packages such as \texttt{crsh/papaja}, \texttt{DF\_network}, and \texttt{swfscMisc}, which were either not available for the R version in the container or required system-level libraries that were not included.
    
    \item \textbf{Failed package installation:} These issues caused \texttt{devtools::install\_local(getwd())} to fail, stopping container creation during the \texttt{repo2docker} build.
\end{itemize}

The remaining 249 projects were successfully containerised, yielding 460 R scripts, and demonstrating the pipeline's ability to automatically generate and build Docker images for a majority of the analysed OSF R projects. A key constraint during the publication step was GitHub’s 100MB per-file size limit, which prevented 23 of these projects (51 scripts) from being pushed to the repository. While this limited their accessibility via Git-based platforms such as Binder, the scripts were still containerised and included in the execution analysis, maintaining the total number of executed scripts at 460.

\subsection{Code execution within containerised environments}
Following successful containerisation, 460 R scripts from 249 OSF projects were executed. Of these, 119 scripts (25.87\%) completed successfully without critical errors, while the remaining 341 scripts (74.13\%) failed—highlighting persistent challenges in computational reproducibility. Among the successful scripts, 51 came from 40 projects (16.06\% of all 249 projects) in which all scripts executed without failure, indicating full project-level reproducibility. The other 68 successful scripts were from 34 projects (13.65\%) that also included at least one failed script, reflecting partial reproducibility. The remaining 175 projects (70.28\%) had no successfully executed scripts.

To analyse and interpret script execution failures, a two-level classification approach was implemented. At the first level, regular expression patterns were used to extract common error messages and map them to initial error types, such as missing objects, function call issues, invalid paths, or package loading problems. In the second level, semantically similar errors were grouped into broader categories.

Errors such as ``object not found,'' ``could not find function,'' and ``unexported object access'' were grouped under \textit{Missing Object or Function} (18.2\%), which also includes references to undefined variables or function calls made without properly loaded packages. Errors involving non-existent file or folder paths, hardcoded directories, or failed attempts to change the working directory were classified as \textit{Invalid File or Directory Path} (19.1\%). Typical cases include ``cannot open file,'' ``failed to search directory,'' or ``directory already exists,''  reflecting file system access issues. Many errors stemmed from scripts referencing missing datasets or using absolute paths, both of which hinder reproducibility in containerised or remote environments.

Package-related issues were among the most frequent failure sources, with \textit{Missing Package} errors accounting for 26.1\% of failed scripts. This category includes unmet dependencies due to missing, outdated, or deprecated packages, or libraries implicitly required but not declared. Separately, errors during installation, such as broken dependencies, compilation failures, or loading issues, were categorized as \textit{Package Installation Failure} (8.2\%), often signaled by messages like unable to install packages,'' lazy loading failed,'' or ``package or namespace load failed.''

System-level problems related to shared object files or display devices were categorized under \textit{Shared Library Load Error} (8.5\%). These include failures such as ``unable to load shared object'' or GUI-dependent functions failing in headless environments (e.g., ``unable to start data viewer''). Direct file access errors, such as failures when trying to read a file, were grouped under \textit{File Read Error} (7.9\%). These typically manifested as errors in reading \texttt{.rds}, \texttt{.csv}, or other files due to missing or incorrect paths.

The \textit{Other Errors} category, accounting for 12\% of all failed scripts, includes less frequent but collectively significant error types. A substantial portion involved \textit{RStudio Environment Errors} and \textit{Compressed File Not Found} errors, often caused by assumptions about interactive sessions (e.g., active RStudio) or attempts to load missing \texttt{.rds} or compressed files. Less common issues included \textit{Syntax or Argument Errors}, \textit{Encoding and String Handling Problems}, \textit{Missing Arguments in \texttt{setwd()}}, and \textit{Data Structure Mismatches}. Though individually rare, these represent a long tail of reproducibility challenges in real-world R scripts.

\begin{figure}[!t]
    \centering
    \includegraphics[width=\linewidth]{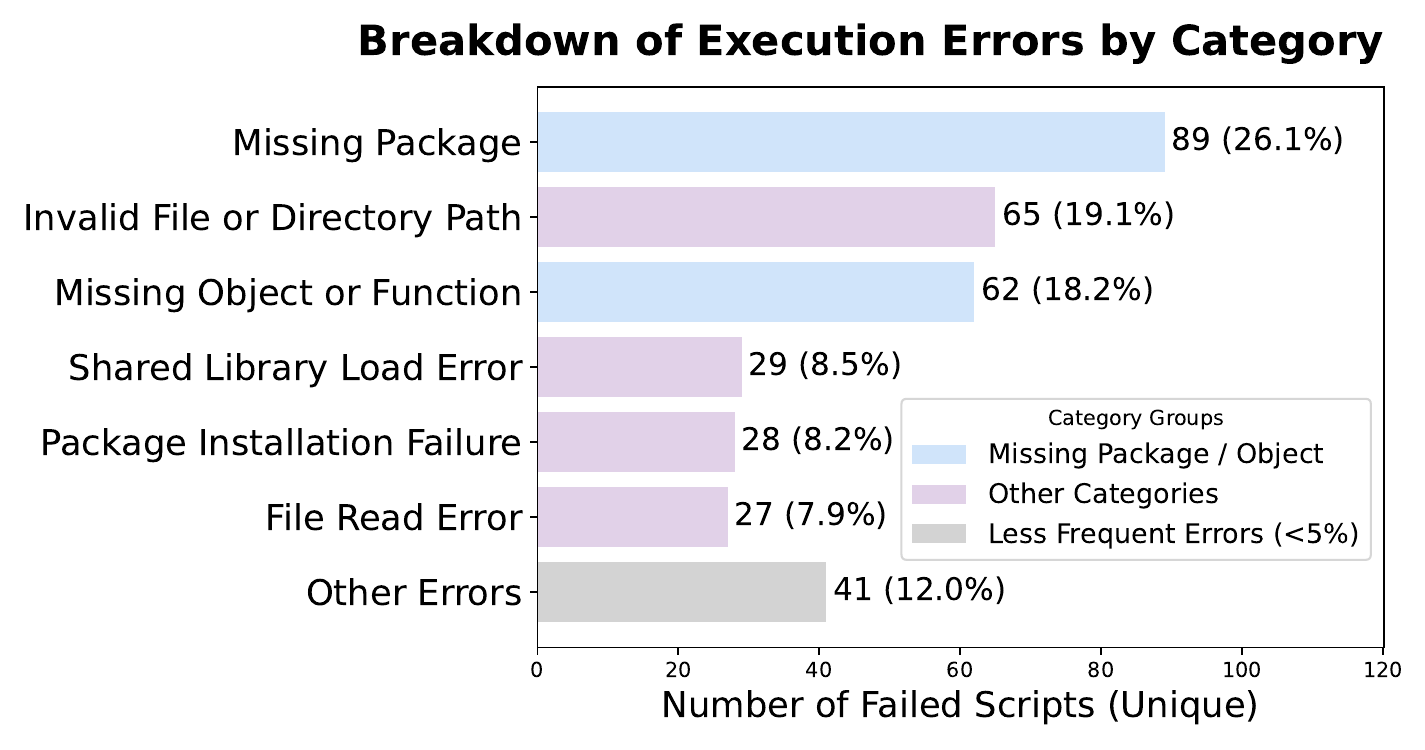}
    \caption{Breakdown of execution errors by category.}
    \label{fig:error_pdf}
\end{figure}

This structured classification enabled the distillation of hundreds of distinct error messages into a concise set of meaningful categories, as visualized in the final breakdown of execution errors (see Figure~\ref{fig:error_pdf}).

\section{Conclusion}
This study examined the reproducibility of statistical R scripts hosted on the Open Science Framework (OSF) using the \texttt{StatCodeSearch} dataset and a new automated container-based pipeline, \texttt{osf-to-binder}. A total of 558 R scripts from 296 OSF projects were analysed, of which 460 scripts were successfully containerised and executed. Execution outcomes were categorized and analysed to uncover prevalent failure patterns. Our findings corroborate earlier results by \citet{hardwicke2020,trisovic2022,pimentel2019} and highlight common barriers to reproducibility, including unresolved dependencies, non-portable file paths, and assumptions about graphical execution environments.

The analysis highlights several best practices to improve the reproducibility of R scripts. First, all required packages should be explicitly declared and sourced from reliable repositories such as CRAN, Bioconductor, or GitHub, with version information when relevant. Providing installation scripts (e.g., \texttt{install.packages()} or \texttt{renv::restore()}) simplifies environment setup. Second, scripts should use relative paths (e.g., \texttt{"./data/data.csv"}) instead of hardcoded absolute paths, and input data should be included or linked via persistent references. Scripts should be verified for required files at runtime and return informative errors if missing. Finally, to support execution in automated or containerised environments, scripts should run non-interactively and avoid GUI-based functions in favor of programmatic alternatives, enabling full execution without manual intervention. 

Cross-domain comparisons suggest that the reproducibility challenges observed in R are not unique. For instance, \citet{samuel2024} analysed 27,271 Jupyter notebooks associated with biomedical publications. Among the 10,388 Python notebooks for which all declared dependencies could be successfully installed and execution was attempted, 1,203 notebooks (11.6\%) ran without any errors. In comparison, our study found that 119 out of 460 R scripts (25.87\%) executed without errors. Although drawn from different research domains and computational ecosystems, both studies point to the persistence of reproducibility barriers even when execution environments are successfully constructed. These findings reinforce the importance of language-agnostic best practices such as explicit dependency declaration, non-interactive design, and environment encapsulation for reproducible research.

However, this work is subject to certain limitations: 
\begin{itemize} 
    \item \textbf{Dependency extraction with \texttt{flowR}}: The \texttt{flowR} static dataflow analyser is limited in detecting all import patterns in R, especially dynamic loading methods (e.g., via \texttt{lapply})\footnote{\url{https://github.com/flowr-analysis/flowr/issues/1259}}. Some required packages may remain undetected, leading to incomplete environment specifications and potential containerisation failures.
    \item \textbf{Single-error reporting:} Execution halts at the first critical error, preventing analysis of subsequent issues in the same script.
    \item \textbf{Limited GitHub compatibility:} Projects with large files cannot be pushed to GitHub without Git LFS, restricting their availability via platforms like Binder, though they remain executable in the pipeline.

\end{itemize}

Future work could enhance dependency detection through dynamic code analysis or AI-assisted parsing. Expanding the pipeline to support multi-error detection and fault-tolerant recovery could offer a fuller picture of reproducibility challenges. Addressing GitHub’s size constraints through automated Git LFS integration or alternative hosting strategies (e.g., Zenodo, OSF storage links) would improve accessibility and enable broader sharing of data-intensive projects. Encouraging broader adoption of OSF registrations and promoting best practices for file versioning at the time of publication (e.g., linking analysis scripts explicitly to registration DOIs) could also strengthen long-term reproducibility. Finally, integrating large language models and intelligent debugging systems may provide promising avenues for reducing execution failures and streamlining reproducibility assessments in computational research.

\vspace{1em} \noindent This work was funded by the German Research Foundation (DFG) under project No. 504226141.

\appendix

\bibliography{aaai25}

\end{document}